\documentclass[aps,]{revtex4}%
\usepackage{amsfonts}
\usepackage{amsmath}
\usepackage{amssymb}
\usepackage{graphicx}%
\begin{document}
\title[Gap in graphene]{Dynamical creation of gap in the monolayer graphene}
\author{Sergey A. Ktitorov}
\affiliation{A.F. Ioffe Institute, St. Petersburg, Russia, and the State Electrotechnical
Institute, St. Petersburg, Russia.}
\author{Xiaoxing Chen}
\affiliation{The State Electrotechnical Institute, St. Petersburg, Russia.}
\keywords{Gross-Neveu model, marginal dimension.}

\begin{abstract}
The zero gap electronic bands in the monolayer graphene are shown to be
unstable relative to the dynamic symmetry violation due to the electron-phonon interaction.

\end{abstract}
\maketitle

%\preprint{ }

%\author{Third Author}

%\pacs{PACS number}

%\volumeyear{year}
%\volumenumber{number}
%\issuenumber{number}
%\eid{identifier}
%\date[Date text]{date}
%\received[Received text]{date}

%\revised[Revised text]{date}

%\accepted[Accepted text]{date}

%\published[Published text]{date}

%\startpage{101}
%\endpage{102}
%\tableofcontents

\section{Introduction}

On the face of it, the massless Dirac spectrum of the monolayer graphene is
well established\cite{neto}. However, there is an evidence that some narrow
gap can really exist \cite{gap}. It is usually traced back to the mutual
displacement of the sublattices. The question arises: what is the nature of
this displacement. One of the possibilities is the effect of structure
defects. Our goal here is to investigate an alternative possibility: a
dynamical symmetry break accompanied by generation of a gap (mass) due to
strong enough electron-phonon interaction. Such phenomenon is well studied in
the superconductivity \cite{schrieff}, Gross-Neveu model of the quantum field
theory \cite{gross-neveu}, and the theory of organic quasi-one-dimensional
conductors \cite{su}. Notice that the system dimension plays a crucial role in
this problem. We are going to show, how and why this phenomenon can take place
in the 2+1 space-time.

\section{Electron-phonon system of graphene}

The Lagrangian of the Dirac electron interacting with the optical phonons
reads%
\begin{equation}
L=\sum_{\mu=1}^{2}\sum_{a=1}^{2}\left(  i\hbar v_{F}\overline{\psi}_{a}%
\gamma_{\mu}\partial_{\mu}\psi_{a}-i\hbar\overline{\psi}_{a}\gamma_{0}%
\partial_{0}\psi_{a}-g\varphi\overline{\psi}_{a}\psi_{a}\right)  +\rho
\omega_{0}^{2}\varphi^{2}/2, \label{lagrange}%
\end{equation}

where $\gamma_{x}=i\sigma_{y},\gamma_{y}=-i\sigma_{x}$ are the Pauli matrices,
$\psi$ is the 2-spinor N-component wave function, $\varphi$ is the scalar
field representing the lattice oscillation normal coordinate corresponding to
the dispersionless optical phonon mode with the frequency spectrum
$\omega\left(  k_{x},k_{y}\right)  =\omega_{0}$, $g$ is the electron-phonon
interaction constant, $v_{F}$ is the electron velocity near the Dirac point.
The number of components N stands for the number of fermion species (points
$K$ and $K\prime$in the Brillouin zone).\ 

\section{Gap equation}

The standard procedure of the fermion integrating away \cite{zinn} from the
derivative functional
\begin{align}
Z  &  =\int D\varphi\exp\left[  -\frac{\rho\omega_{0}^{2}}{2}\int d^{2}%
x\int_{0}^{\beta}d\tau\varphi^{2}\left(  x,\tau\right)  \right]
\times\nonumber\\
&  \int D\overline{\psi}D\psi\exp\left[  -\sum_{a=1}^{2}\int_{0}^{\beta}%
d\tau\int d^{2}x\overline{\psi}_{a}\left(  -%
%TCIMACRO{\dsum \limits_{\mu=1}^{2}}%
%BeginExpansion
{\displaystyle\sum\limits_{\mu=1}^{2}}
%EndExpansion
v_{F}\gamma^{\mu}\widehat{p}_{^{\mu}}-i\hbar\gamma_{0}\partial_{0}%
-g\varphi\right)  \psi_{a}\right]  \label{partition}%
\end{align}
leads to the effective action for the classical field $\varphi:$%

\begin{equation}
S=-\frac{\rho\omega_{0}^{2}}{2}\int d^{2}xd\varphi^{2}+tr\int_{0}^{\beta}%
d\tau\int d^{2}x\left[  \ln\left(
%TCIMACRO{\dsum \limits_{\mu=1}^{2}}%
%BeginExpansion
{\displaystyle\sum\limits_{\mu=1}^{2}}
%EndExpansion
v_{F}\gamma^{\mu}\widehat{p}_{^{\mu}}+i\hbar\gamma_{0}\partial_{0}%
+g\varphi\right)  \right]  . \label{effective}%
\end{equation}

The stationary phase condition%
\begin{equation}
\frac{\delta\widetilde{S}}{\delta\phi}=0 \label{statphase}%
\end{equation}

gives us the self-consistency equation%

\begin{equation}
\phi=\frac{1}{\left(  2\pi\right)  ^{2}}\widetilde{T}\sum_{s=-\infty}^{\infty
}\int^{\widetilde{\Lambda}}d^{2}\kappa\frac{\widetilde{g}\phi}{-\omega_{s}%
^{2}+\kappa^{2}+\widetilde{g}^{2}\phi^{2}} \label{selfconsistgeneral}%
\end{equation}

We have introduced the dimensionless variables and parameters:%

\[
\overline{\Psi}=\overline{\psi}a,\text{ \ }\Psi=\psi a,\text{ \ }\kappa_{\mu
}=k_{\mu}a,\text{ \ }\phi=\frac{\varphi}{a},\text{ \ }\Omega^{2}=\frac
{M\omega_{0}^{2}a}{v_{F}\hbar},\text{ \ }\widetilde{g}=\frac{g}{v_{F}%
\hbar\Omega},\text{ \ }\widetilde{T}=Ta/\hbar v_{F}.
\]

Summing up over the Matsubara frequencies
\begin{equation}
\omega_{s}=i(2s+1)\pi\widetilde{T} \label{matsubara}%
\end{equation}

we obtain%
\begin{equation}
1=\frac{\widehat{g}}{\left(  2\pi\right)  ^{2}}\int^{\widetilde{\Lambda}%
}d\kappa\frac{\kappa}{2\sqrt{\kappa^{2}+\widetilde{g}^{2}\phi^{2}}}\tanh
\frac{\sqrt{\kappa^{2}+\widetilde{g}^{2}\phi^{2}}}{2\widetilde{T}},
\label{dimlessselfconsist}%
\end{equation}

where $\widetilde{\Lambda}=\Lambda a$ is the dimensionless UV cut-off,
$\widetilde{\Lambda}$ $=\pi.$ The following self-consistency equation for
$\widetilde{T}=0$ follows:%
\begin{equation}
1=\frac{\widetilde{g}}{\left(  2\pi\right)  ^{2}}\left[  \sqrt{\widetilde
{g}^{2}\phi^{2}+\widetilde{\Lambda}^{2}}-\widetilde{g}\phi\right]  .
\label{zerotemp}%
\end{equation}

In the case of $\widetilde{T}\neq0$ we have the equation%
\begin{equation}
\frac{\left(  2\pi\right)  ^{2}}{\widehat{g}}=\widetilde{T}\ln\left[
\cosh\left(  \frac{\sqrt{\widetilde{\Lambda}^{2}+\widehat{g}^{2}\phi^{2}}%
}{2\widetilde{T}}\right)  \right]  -\widetilde{T}\ln\left[  \cosh\left(
\frac{\widehat{g}\phi}{2\widetilde{T}}\right)  \right]  \label{finitetemper}%
\end{equation}

Graphical solution of these equations is illustrated in Fig. 1.

The UV cut-off is necessary in this formula for $d>1.$This formula is
asymptotically exact at the limit of $N\rightarrow\infty;$ otherwise it can be
considered as just the mean field theory result. In the one-dimensional case,
it gives for the dynamically generated mass the well known result
$M\propto\varphi_{c}=\Delta\exp\left(  -\frac{1}{Ng}\right)  ,$where cut-off
$\Delta$ stands for the electronic band width. This formula can be useful for
the case of the carbone nano-tubes, which can be considered as a dimensionally
reduced graphene. The gap is not zero at arbitrarily weak electron-phonon
interaction in this case. The situation is different in the case of the
graphene. A threshold magnitude of the interaction constant does exist in this
case:%
\begin{equation}
Ng_{c}=\left[  \int\frac{d^{3}p}{\left(  2\pi\right)  ^{3}\left(
p^{2}\right)  }\right]  ^{-1}. \label{threshold}%
\end{equation}
The gap will be open at $\widehat{g}$ $>\widehat{g}_{cr}=\frac{\left(
2\pi\right)  ^{2}}{\widetilde{\Lambda}}:$%
\begin{equation}
M\propto\Delta\left(  g-g_{c}\right)  \label{mass}%
\end{equation}
Putting $\phi$ to zero in Eq. (\ref{finitetemper}), we obtain the phase
boundary equation $F(g,T)=0$, separating the massive and zeromass phases (see
Fig. 2).

In conclusion, we have shown that a narrow gap can be dynamically created in
the monolayered graphene in the case of the strong enough electron-phonon interaction.

\bigskip

\section{Figure captions}

Fig. 1. Graphical solution of the equation Eq. (\ref{finitetemper}). Thick
(black), medium (red), and thin (green) curves correspond to increasing temperatures.

\bigskip

Fig. 2. Phase boundary $F(g,T)=0.$

\end{document}